\documentclass{aa}
\usepackage{graphicx}
\begin{document}

   \title{The Galaxy Populations of Double Cluster RX J1053.7+5735 at z=1.13}

   \author{Yasuhiro Hashimoto\inst{1}
          J. Patrick Henry \inst{1,2}
          %G\"unther Hasinger\inst{1}
          G. Hasinger\inst{1}
          G. Szokoly \inst{1}
          M. Schmidt \inst{3}
          \and
          %more\inst{1}
          I. Lehmann \inst{1}
          }

    \institute{Max-Planck-Institut f\"ur extraterrestrische Physik,
              Giessenbachstrasse
              D-85748 Garching, Germany
          \and
              Institute for Astronomy, University of Hawaii, 2680 Woodlawn Drive, Honolulu, Hawaii 96822, USA
          \and 
              Palomar Observatory, California Institute of Technology, MS 320-47 Pasadena, CA 91125, USA
          }

   \offprints{Y.\ Hashimoto,\\ e-mail: hashimot@mpe.mpg.de}

   \date{Received ; accepted}

   \authorrunning{Hashimoto et al.}
   \titlerunning{Galaxies in RX J1053.7+5735}

%\begin{abstract}
\abstract{
We present a study of the galaxy population in the cluster RX J1053.7+5735,
one of the most distant 
X-ray selected clusters of galaxies, which also shows an unusual 
double-lobed X-ray morphology, 
indicative of a possible equal-mass cluster merger.
The cluster was discovered during 
the {\it ROSAT} deep pointings in the direction of the Lockman Hole.
Using Keck-DEIMOS
spectroscopic observations of galaxies
in the 2\farcm0 $\times$ 1\farcm5 region surrounding
RX J1053.7+5735, we
secured redshifts for six
galaxies
in the range
1.129 $<$ $z$ $<$ 1.139, with a mean redshift $<z>$=1.134.
The mean redshift agrees well with the
cluster {\em X-ray} redshift previously estimated
from
the cluster X-ray Fe-K line,
confirming the presence of a cluster at $z$$\sim$1.135.
Galaxies with concordant redshifts
are located in both eastern and western sub-clusters of
the double cluster structure,
indicating that
both sub-clusters are at similar redshifts.
This result is also consistent with a previous 
claim that both eastern and western X-ray lobes
have similar X-ray redshifts.
Based on their 
separation of $\sim$ 250 kpc/h,
these results support the interpretation that
RX J1053.7+5735 is an equal-mass cluster merger taken
place at $z$ $\sim$ 1, although
further direct evidence for dynamical state of the cluster is
needed for a more definitive statement about
the cluster merging state.
The six galaxies have a line-of-sight velocity
dispersion $\Delta$$v$ $\sim$ 650 km s$^{-1}$. 
All six galaxies show clear absorption features of CaII H \& K, and
several Balmer lines, typical of early galaxies at the present
epoch, in agreement with their $I-K$ colors.
A color-magnitude diagram, constructed from
deep optical/NIR observations of the RX J1053.7+5735 field,
shows a clear red color sequence.
There is an indication that the red sequence
in RX J1053.7+5735 lies $\sim$ 0.3 to the blue of the Coma line,
qualitatively consistent with previous studies investigating
other clusters at z$\sim$ 1.

%\end{abstract}
}

   \maketitle

\section{Introduction}

The identification and study of distant galaxy clusters is
of great interest in current astronomical research.
As the most massive collapsed objects in the universe,
clusters provide a sensitive probe of the formation and evolution structure
 in the universe 
(Eke, Cole, \& Frenk 1996; Bahcall, Fan, \& Cen 1997). 
The presence of clusters or large-scale structure at and
beyond z = 1 can constrain scenarios for bottom-up structure formation and fundamental
cosmological parameters 
by requiring
cluster-level collapse earlier than
that epoch.
Another equally important use of galaxy clusters lies in
studying the formation and evolution of galaxy populations. 
The study of the spectral and photometric properties of cluster
galaxies at large redshifts provides a powerful means of discriminating
between scenarios for the formation of  elliptical galaxies.
Predictions of the color-magnitude distribution of cluster early-types galaxies based on
hierarchical models of galaxy formation
(Kauffmann 1996; Kauffmann \& Charlot 1998) differ at z $>$ 1 from
models in which ellipticals formed at high-z in a monolithic collapse
(Eggen, Lynden-Bell, \& Sandage 1962).
Meanwhile, study of the individual galaxies in high-z 
clusters can also provide
information on the cluster environment, merging and star formation history
in the cluster galaxies, and samples of cluster ellipticals and spirals which could be
compared with field galaxies at the high redshift and cluster galaxies at low redshift.

The cluster RX J1053.7+5735,
which shows an unusual double-lobed
X-ray morphology, 
was discovered during 
deep (1.31 Msec) {\it ROSAT} HRI pointings
(\cite{Hget98}), 
in the direction of the ``Lockman Hole",
a line of sight with exceptionally low HI column density
(Lockman et al. 1986).
The angular size of the source is 1.7 $\times$ 0.7 arcmin$^2$
and 
two lobes are approximately 1 arcmin apart.
The total X-ray flux of the entire source 
in the 0.5-2.0 keV band 
is 2 $\times$ 10$^{-14}$ erg cm$^{-2}$ s$^{-1}$
(\cite{Hget98b}).
The subsequent deep optical/NIR imaging follow-ups (V $<$ 26.5, R $<$ 25, I $<$ 25,
K $<$ 20.5) with LRIS and NIRC on Keck,
and the Calar Alto Omega Prime camera
revealed a bright 7 arcsecond-long arc with a
magnitude of R=21.4 as well as
an overdensity of galaxies in both X-ray lobes
(e.g. \cite{Thet01}).
Further Keck LRIS/NIRSPEC spectroscopic observations on the bright arc
and one of the brightest galaxies near the arc showed that the former
is a lensed galaxy at a redshift z = 2.57
while the latter may be at a redshift of z = 1.263 (\cite{Thet01}).  
Deep VRIzJHK photometry data 
produced concordant
photometric redshifts for more than 10 objects
at redshift of z $\sim$ 1.1-1.3, confirming
that at least the eastern lobe is a massive cluster at high redshift.
The improbability of chance alignment and similarity of colors for the galaxies in the
two X-ray lobes were consistent with the western lobe also being at z $\sim$ 
1.1-1.3 (\cite{Thet01}).

The X-ray data of the XMM-Newton (XMM) observation (with a total 
effective exposure time
$\sim$ 100 ks)
performed during the PV phase for this cluster 
were analyzed 
(Hashimoto et al. 2002), yielding   
a best-fit temperature of 4.9 $^{+1.5}_{-0.9}$ keV,
while the metallicity was poorly constrained 
with an upper limit on the iron abundance of 0.62$Z_{\odot}$.
Hashimoto et al. (2004), using even deeper ($\sim$ 700 ks) XMM observations,
detected the Fe K line in the cluster X-ray emission and obtained a strong constraint on
cluster metallicity, which is difficult to achieve for clusters at z $>$ 1.
The best-fit abundance is  0.46  $^{+0.11}_{-0.07}$ times the solar value.
Comparison with other metallicity measurements of nearby and
distant clusters showed that there was little evolution
in the ICM metallicity from z $\sim$ 1 to the present.
The Fe line emission also allowed them to directly estimate the redshift of
diffuse gas, with a value  z = 1.14 $^{+0.01}_{-0.01}$.
This is one of the first clusters whose
X-ray redshift is directly measured prior to the secure knowledge of
cluster redshift by optical/NIR  spectroscopy.
Hashimoto et al. (2004) could also estimate the X-ray redshift separately for each
of the two lobes in the double-lobed structure, and the result was
consistent with the
two lobes being part of one cluster system
at the same redshift.

Here we report on the optical imaging and spectroscopic follow-up 
studies of the cluster RX J1053.7+5735 based mainly on our deep observations with 8m-class telescopes.
The paper is organized as follows. In Sec. 2, we briefly
describe the observation and data reduction.
In Sec. 3, we present imaging and spectroscopic analysis of the
cluster RX J1053.7+5735.
Sec. 4 summarizes out results.
Throughout the paper, we use $H_{o}$ = 65 km s$^{-1}$ Mpc$^{-1}$,
$\Omega_{m}$=0.3, and $\Omega_{\Lambda}$=0.7 

\section{Observations and Data}
\subsection{Optical and IR Imaging}
Deep $R$, $I$, \& $Z$ band images of the region
surrounding
the cluster 
RXJ 1053.7+5735 were obtained 
as a part of IfA Deep Survey (Barris et al. 2004)
using 
the Suprime-Cam 
(Miyazaki et al. 1998)
on the Subaru telescope
over ten nights from 2001 November through 2002 April.
The camera covers a 34' $\times$ 27' field of view with a
pixel scale of 0\farcs20.
The $Z$ filter at Subaru has an effective wavelength of 9195 \AA\ and
FWHM of 1410 \AA\ (Fukugita et al. 1996), while $R$ and $I$ filters are Cousins
$R$ and $I$.
The data were taken under various seeing conditions, 
and we used only
images with less than $\sim$ 0\farcs8 seeing.
Total effective exposure time after this filtering
is 1680s, 2150s, 2640s in $R$, $I$, \& $Z$ band, respectively.
The photometry is calibrated to Vega system using Landolt standards (Landolt 1992).
For further details of the Subaru data, please see Barris et al. (2004)

The infrared $K$ ($K'$) images were obtained in 1994 Nov. and 2002 Jan. at the Calar Alto 3.5m telescope
with the Omega-Prime infrared camera, providing a 6\farcm8 
field of view with 0\farcs39 pixels.
The flux scale was calibrated using five UKIRT faint standards
(Casali \& Hawarden 1992) obtained during the 2002 January run.
The data were taken using dithered motion with amplitude of 10".
The total integration time and seeing of
the resulting image are 155 min and 1\farcs4.

\begin{figure}
 \resizebox{\hsize}{!}{\includegraphics[clip]{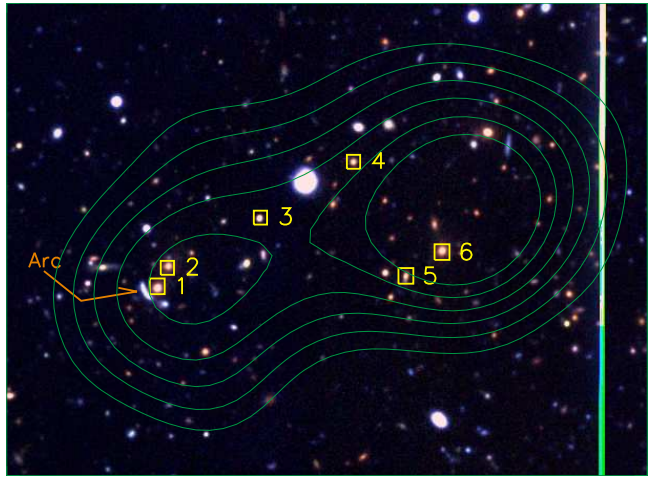}}
 \caption{
The color image of
2\farcm0 $\times$ 1\farcm5 area around the cluster RX J1053.7+5735.
The image is composed using $R$, $I$, and $Z$ images taken with Subaru Suprime-
Cam.
North is up and East is left.
Spectroscopically confirmed members (boxes) and their ids are shown.
The contour is based on the X-ray image obtained by XMM,
combining all events in the 0.2-8 keV band
from three
(pn, MOS1, \& MOS2) cameras.
The lowest contour is 1.9 counts arcsec$^{-2}$ and
the contour interval is 0.2 counts arcsec$^{-2}$.
}
\label{FigTemp}
\end{figure}

\subsection{Optical Spectroscopy}
Spectroscopic observations of cluster galaxies were obtained 
as a part of optical follow-up program of $\sim$ 140 X-ray sources 
detected in the Lockman Hole,
using
the Deep Imaging Multi-Object Spectrograph (DEIMOS; Faber et al. 2003)
on the Keck II telescope.
Four masks covering a total of 16 $\times$ 20 arcmin$^2$ area
were used over a four night observing-run in 2004 February,
resulting in a total integration time of $\sim$ 8 hours for each object.
Using one mask covering the cluster region, 
we assigned four tilted slits to cover eight potential cluster members.
No fringing is present in the red for DEIMOS, 
allowing us to use tilted slits, because 
there is no need to construct
a fringe frame
using dithered exposures.
Cluster galaxies were assigned slits based mainly on their $I-K$ color
and $K$ magnitude.
We used 
the 600 lines mm$^{-1}$ grating  blazed at 7150 \AA,
covering 4550-9700 \AA.
The dispersion is  0.65 \AA/pixel. 
Spectral resolution depends on the angle of the tilted slit, but
it is typically $\sim$ 6 \AA.
The slit mask data were separated into individual slitlet
spectra and then treated as 
standard long slit spectral data.
The integration is split into a series of 1800s exposures. 
The exposures for each slitlet were reduced separately,  
then co-added.
Wavelength calibration of the spectra was obtained from
NeXeCdZnHg arc lamp exposures taken in the same night.
A flux calibration was obtained from long-slit
observations of the standard stars HZ44 and LTT6028.

\begin{figure}
 \resizebox{\hsize}{!}{\includegraphics[clip,angle=90]{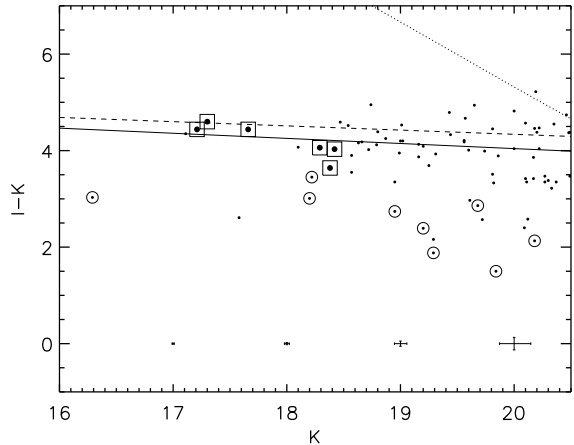}}
 \caption{
Color-magnitude diagram for galaxies inside a  
2\farcm0 $\times$ 1\farcm5
field surrounding the cluster RX J1053.7+5735.
Colors are based on $I$ band Subaru Suprime-Cam image, and $K$ band 
Calar Alto Omega-Prime image.
Spectroscopically confirmed members are marked by squares, while spectroscopically known foreground galaxies are marked by
circles.
The error bars near the bottom of the plot indicate 
the size of 1 $\sigma$ error.
The dotted line marks the 5 $\sigma$ limit in $I-K$ color. 
The dashed line shows the CM relation of Coma
transformed to the cluster redshift assuming no evolution.
The solid line is a linear fit to the CM relation of galaxies with
$K$ $<$ 20, $I-K$ $>$ 3, and $I-K$ $<$ 6, excluding the known foreground galaxies.
}
\label{FigTemp}
\end{figure}

\section{RESULTS}
\subsection{Photometric Properties}
Figure 1 shows 
a color image of 
a 2\farcm0 $\times$ 1\farcm5 area around 
the cluster RX J1053.7+5735.
The image is composed using $R$, $I$, and $Z$ band images taken with Subaru Suprime-Cam.
Spectroscopically confirmed members (boxes) and their ids, as well as a position of the gravitational arc, are indicated.
The contours are from the X-ray image obtained by XMM,
combining all events in the 0.2-8 keV band
from three
(pn, MOS1, \& MOS2) cameras.
The lowest contour is 1.9 counts arcsec$^{-2}$ and
the contour interval is 0.2 counts arcsec$^{-2}$.
A catalog of objects in each optical/IR band  was obtained using
SExtractor(Bertin \& Arnouts 1996). 
The photometry was obtained using a fixed 2'' aperture.
Objects were detected with the requirement that
an object area of 2 arcsec$^2$
must be 1.5$\sigma$ above the background.
Photometric properties of spectroscopically confirmed
galaxies, as well as their spectroscopic redshifts are
shown in the table 1.

Figure 2 shows
a color-magnitude diagram for all objects in a
2\farcm0 $\times$ 1\farcm5
field surrounding the cluster RX J1053.7+5735.
Colors are based on $I$ image taken with Subaru Suprime-Cam, and 
$K$ image taken with Calar Alto Omega-Prime camera.
Spectroscopically confirmed members are marked by squares, while 
spectroscopically known foreground galaxies 
at z=0.05-0.78 from Hasinger et al. (1998b) 
are marked by
circles.
The error bars near the bottom of the plot 
indicate the size of 1 $\sigma$ error.
The dotted line marks the 5 $\sigma$ limit in 
$I-K$ color. 
The dashed line shows the CM relation of Coma
(Stanford et al. 1998)
transformed to the cluster redshift assuming no evolution.
The solid line is a linear fit,
$(I-K)=(4.201\pm0.069)-(0.106\pm0.086)(K-18.5)$, 
to the CM relation of galaxies with
$K$ $<$ 20, $I-K$ $>$ 3, and $I-K$ $<$ 6, excluding the known foreground galaxies.
There is an indication that the red sequence
in RX J1053.7+5735 lies $\sim$ 0.3 to the blue of the Coma line,
qualitatively consistent with previous studies investigating
other clusters at z$\sim$ 1
using optical-NIR color and, unlike our study,
pre-selection of early-type galaxies
by their morphologies (e.g. Stanford et al. 1998; van Dokkum et al. 2001; 
Stanford et al. 2002; Holden et al. 2004).
There is also an indication that
the fitted slope may be similar with the Coma line,
consistent with some high redshift studies
(e.g. Blakeslee et al. 2003; Lidman et al. 2004), while
contradicting studies reporting the slope evolution
(e.g. van Dokkum et al. 2001; Stanford et al. 2002). 
The caution has to be exercised, however, in interpreting our CM diagram,
because of  the limitation of
the data, and due to the lack of morphological pre-selection.

\begin{table}
\scriptsize
\caption[]{Properties of Galaxies in RX J1053.7+5735}
%\begin{center}
%\begin{tabular}{lrrrl}
%\begin{tabular}{lrrrrrl}
\begin{tabular}{lccccrcc}
\hline
\hline
\noalign{\smallskip}
ID       & R.A.   & Decl. & $R$ &  $K$ &  $R$-$Z$ & $I$-$K$ & Redshift \\
         & (10:)     & (+57:)      & (mag) &  (mag) &          &     &   \\
%         &        &       & (mag) &  (mag) &          &     &   \\
\noalign{\smallskip}
\hline
\noalign{\smallskip}
%1 & 10:53:46.88 & +57:35:10.39 &  23.08 &  17.21 & -0.19 & 4.44 & 1.130\\
%2 & 10:53:46.63 & +57:35:14.56 &  23.68 &  17.66 & -0.08 & 4.44 & 1.129\\
%3 & 10:53:44.47 & +57:35:23.79 &  23.55 &  18.38 & -0.15 & 3.64 & 1.139\\
%4 & 10:53:42.26 & +57:35:34.67 &  23.88 &  18.29 & -0.16 & 4.06 & 1.139\\
%5 & 10:53:41.00 & +57:35:12.69 &  23.86 &  18.42 & -0.32 & 4.03  & 1.130\\
%6 & 10:53:40.14  & +57:35:17.77 & 23.48 & 17.30 & 0.00 & 4.6  & 1.135\\
1 & 53:46.88 & 35:10.39 &  23.08 &  17.21 & -0.19 & 4.44 & 1.130\\
2 & 53:46.63 & 35:14.56 &  23.68 &  17.66 & -0.08 & 4.44 & 1.129\\
3 & 53:44.47 & 35:23.79 &  23.55 &  18.38 & -0.15 & 3.64 & 1.139\\
4 & 53:42.26 & 35:34.67 &  23.88 &  18.29 & -0.16 & 4.06 & 1.139\\
5 & 53:41.00 & 35:12.69 &  23.86 &  18.42 & -0.32 & 4.03  & 1.130\\
6 & 53:40.14 & 35:17.77 & 23.48 & 17.30 &  0.00 & 4.60  & 1.135\\
\noalign{\smallskip}
\noalign{\smallskip}
\hline
\end{tabular}
\end{table}

\begin{figure}
 \resizebox{\hsize}{!}{\includegraphics[bb=189 100 600 650,clip,angle=90]{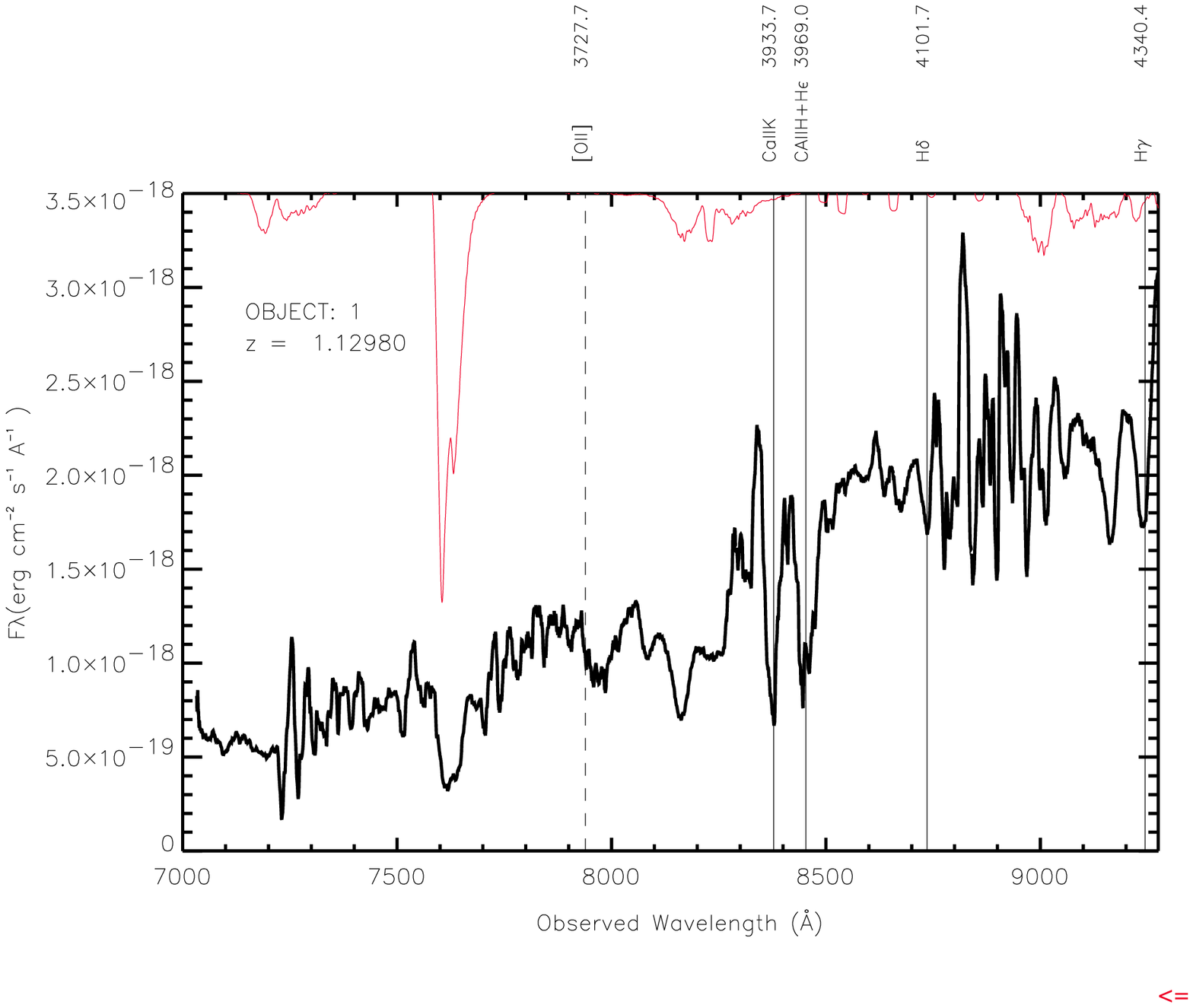}}
 \resizebox{\hsize}{!}{\includegraphics[bb=189 100 487 650,clip,angle=90]{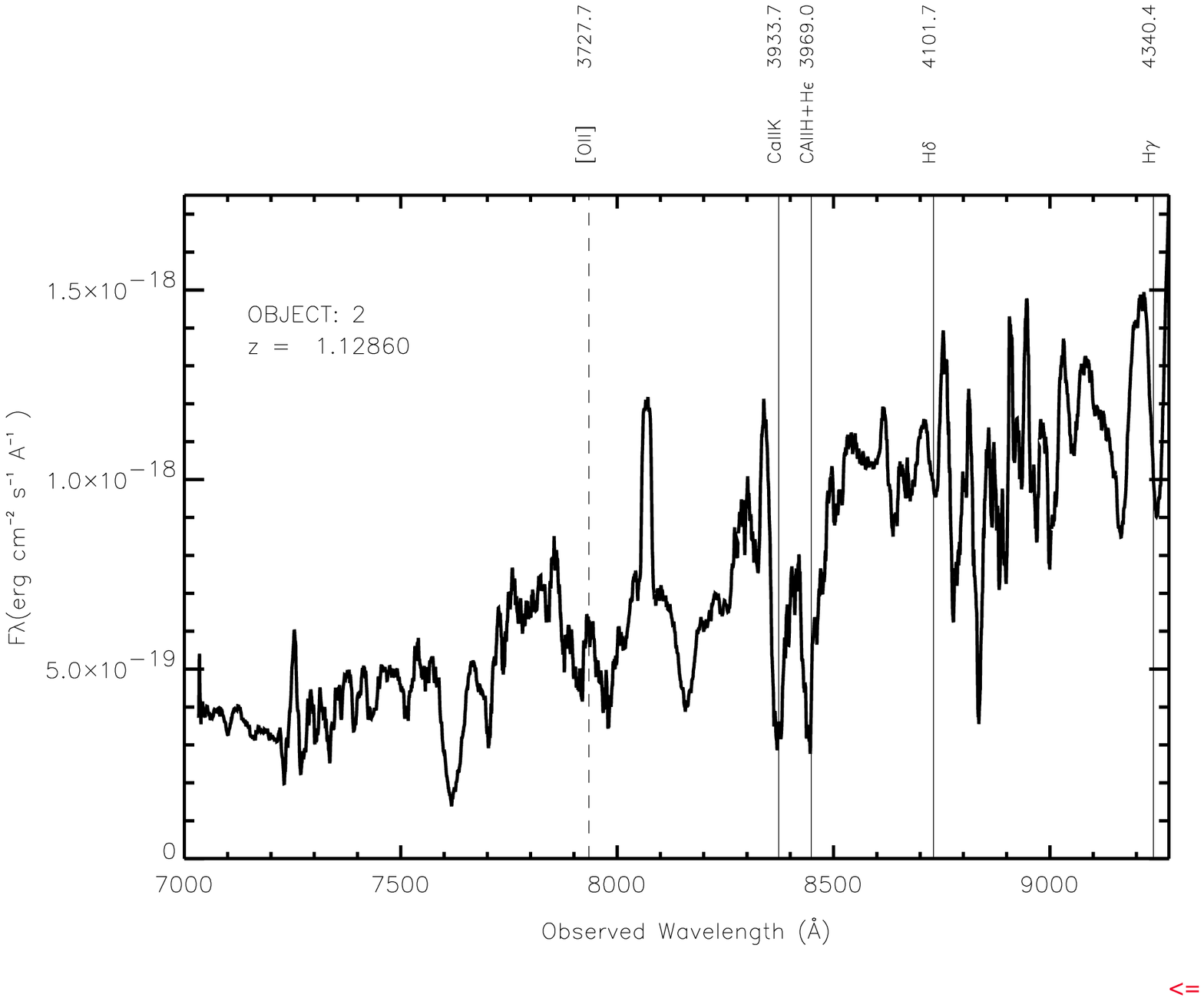}}
 \resizebox{\hsize}{!}{\includegraphics[bb=150 100 487 650,clip,angle=90]{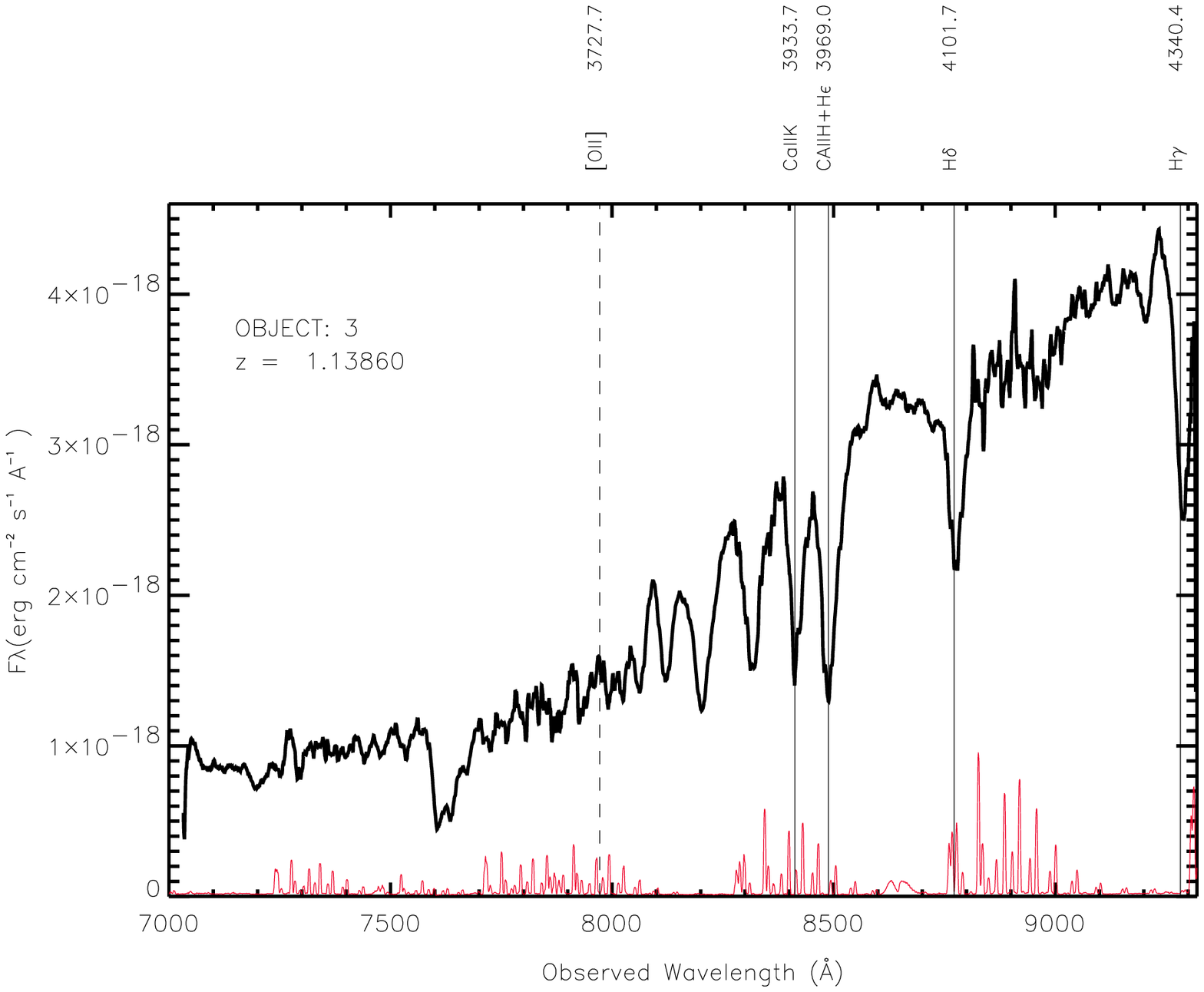}}
 \caption{Optical spectra of eastern members obtained using DEIMOS mounted on Keck telescope.
  Sky emission and absorption  spectra are shown at the bottom and the top of the figure, respectively,
with arbitrary flux levels.
Positions of major spectra features are marked by vertical lines. Solid vertical lines represent features
used for redshift determination, while dashed lines are plotted just as references.
}
\label{FigTemp}
\end{figure}

\subsection{Spectral Properties}
 Figure 3 and figure 4 show optical spectra of cluster members in the
eastern and western subclusters, respectively.
Sky emission and absorption spectra are shown at the bottom and the top of the figure, respectively, with arbitrary flux levels. 
Subtraction of sky {\em emission} lines
is relatively good, considering faint magnitude of the objects.
The spectra of object 1 and 2 show slightly lower sky-subtraction quality,
because they are both observed through a highly tilted slit. 
Meanwhile, some telluric (i.e. sky absorption) features are eminent,
due to the lack of multiplicative sky correction, which is not practical
for faint galaxy spectra.
Positions of major spectral features are marked by vertical lines. Solid vertical lines represent features
used for redshift determination, while dashed lines are plotted just as references.
Major spectral absorption features, such as Ca II H \& K and several Balmer lines were typically used for the determination.
We avoided using 
Mg I $\lambda$3830 absorption feature because of its proximity to
the telluric lines.
Redshifts were calculated by 
centering these major spectral features,
and then fitting a redshift solution using IRAF $rvidlines$ task.
Among eight candidate galaxies, 
six galaxies were found to have redshifts in the range
1.129 $<$ $z$ $<$ 1.139, with a mean redshift $<z>$=1.134 and a 
line-of-sight velocity dispersion $\Delta$$v$ $\sim$ 650 km s$^{-1}$.
The mean redshift agrees well  with the cluster redshift 
derived from
the X-ray spectroscopy by Hashimoto et al. (2004). 
The velocity dispersion may be consistent with 
that expected from cluster X-ray temperature,
although caution has to be exercised in interpreting the
result because of the uncertainty associated with the dynamical status
of the cluster.
All six secure members show clear absorption features of CaII H \& K, and several Balmer lines, typical of early galaxies in the present 
epoch, and consistent with their $I-K$ colors.
An emission like feature at $\sim$8100\AA\ of the object 2 is
an artifact caused by cosmic ray which we are unable to completely 
remove because of its location on a major sky emission line.
One additional galaxy in the eastern lobe shows a single emission-like feature
in otherwise featureless spectrum. If we interpret this as a real 
[OII]$\lambda$3727 emission, this may yield a seventh member galaxy at 
redshift of z=1.122. 
To be conservative, however, we decided to exclude this galaxy 
from the cluster membership, because its redshift was estimated by  
an insecure single feature.
For the last (eighth) galaxy we observed, we are unable to obtain a redshift
due to the lack of any spectral features.

\begin{figure}
 \resizebox{\hsize}{!}{\includegraphics[bb=189 100 600 650,clip,angle=90]{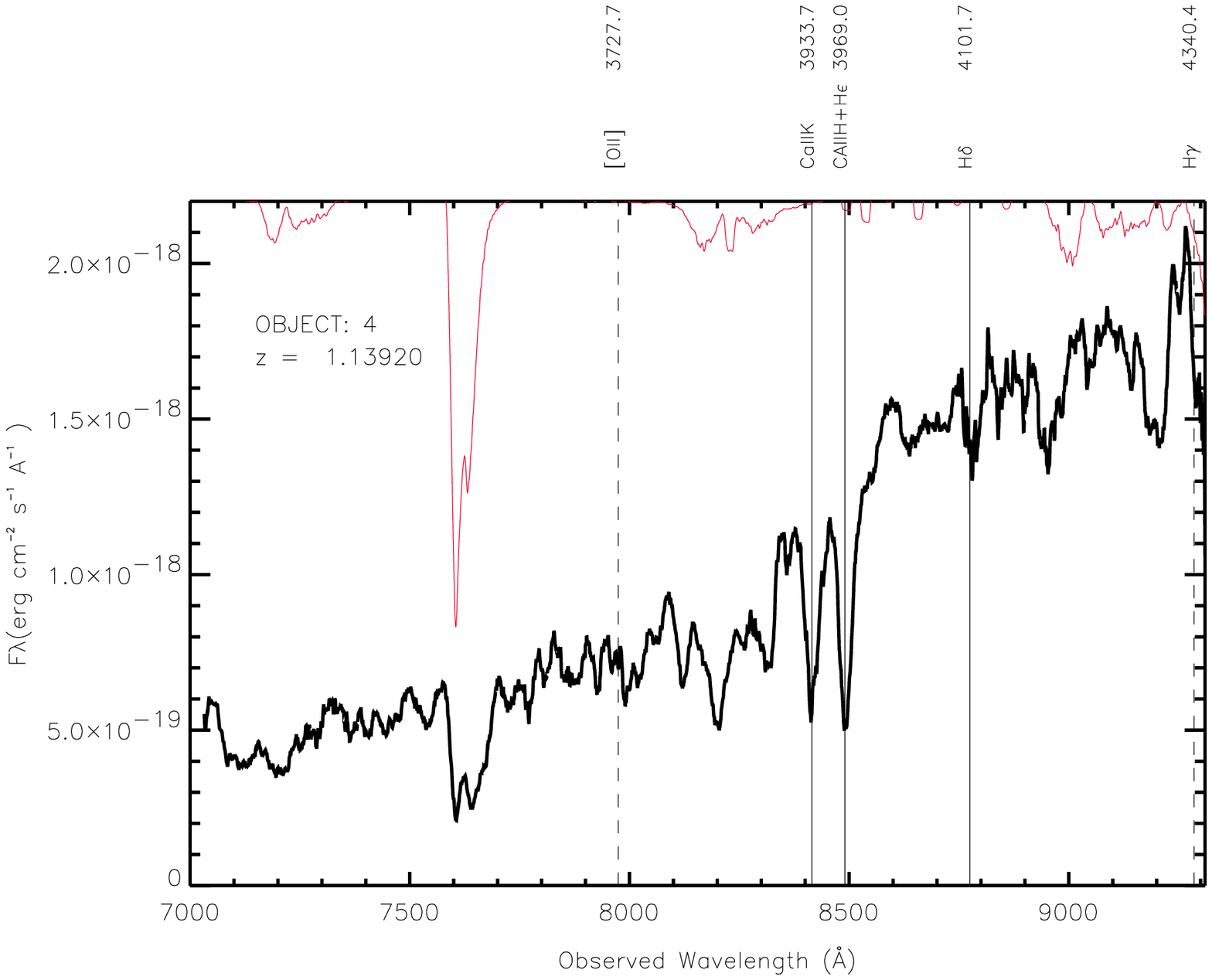}}
 \resizebox{\hsize}{!}{\includegraphics[bb=189 100 487 650,clip,angle=90]{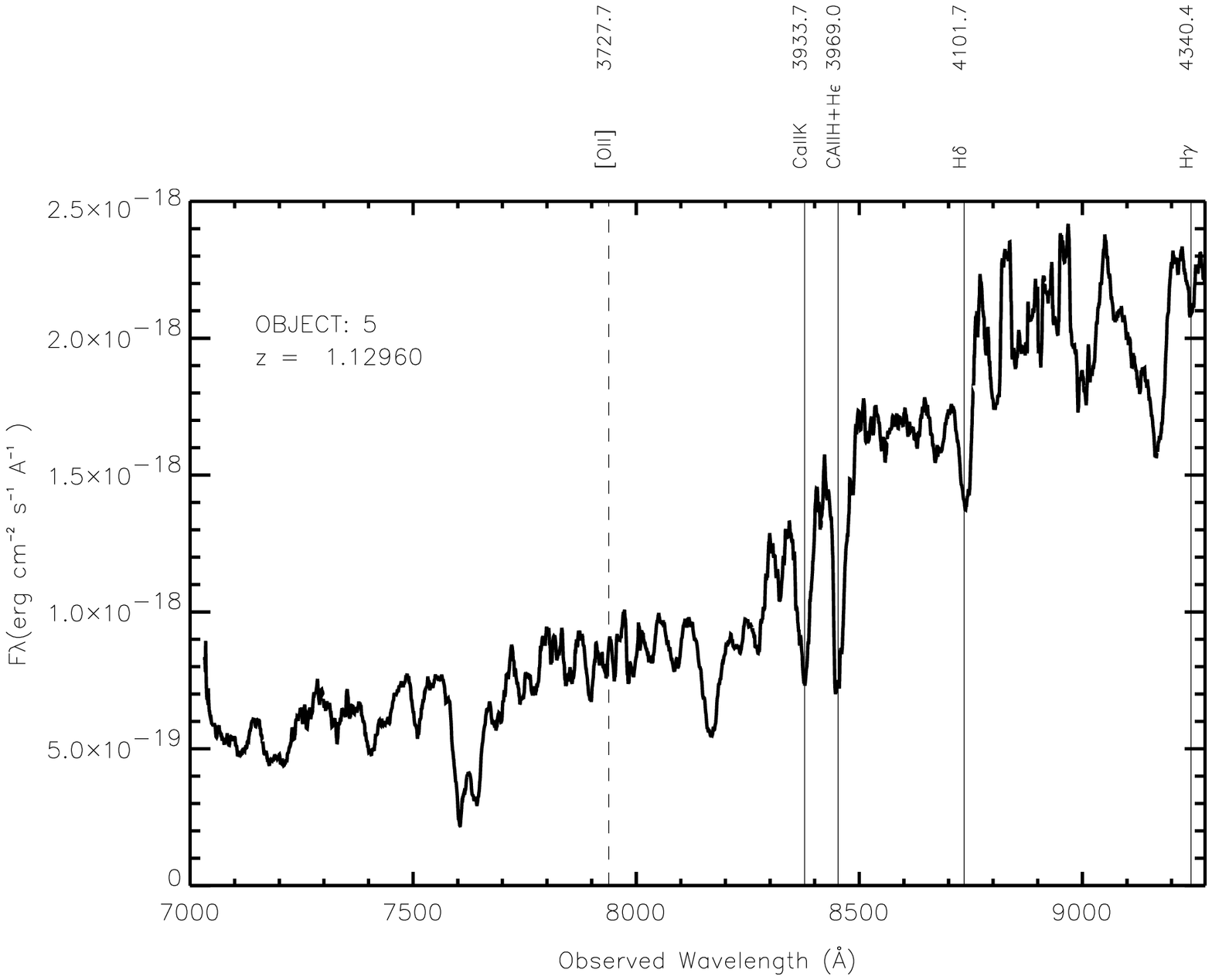}}
 \resizebox{\hsize}{!}{\includegraphics[bb=150 100 487 650,clip,angle=90]{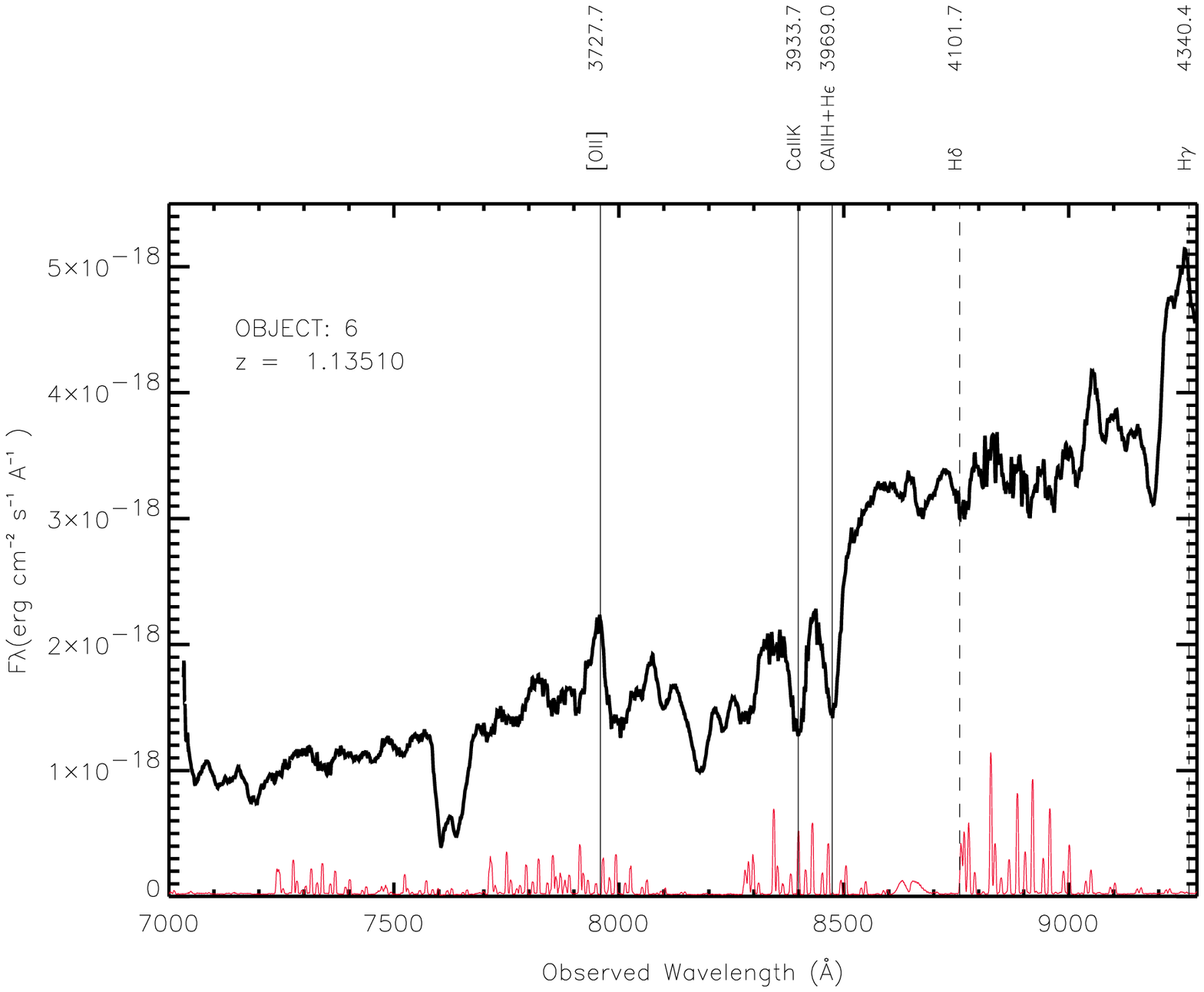}}
 \caption{Optical spectra of western members obtained using DEIMOS.
 Symbols are the same as Fig. 3.
}
\label{FigTemp}
\end{figure}

\section{Summary}
Using Keck-DEIMOS
spectroscopic observations of galaxies
in the 2\farcm0 $\times$ 1\farcm5 region surrounding 
the X-ray detected cluster candidate 
RX J1053.7+5735, we secured redshifts for six 
galaxies 
in the range
1.129 $<$ $z$ $<$ 1.139, with a mean redshift $<z>$=1.134.
The mean redshift agrees well with the  
cluster X-ray redshift previously estimated 
by Hashimoto et al. (2004) using the cluster X-ray Fe-K line,
confirming the presence of a cluster at $z$$\sim$1.135.
Galaxy ID 1  shows a clear redshift z=1.13, being
consistent with previous photometric redshift, but inconsistent
with NIR spectroscopic redshift estimated by Thompson et al. (2001).
Galaxies with concordant redshifts
are located in both eastern and western sub-clusters of the 
double cluster structure,
indicating that
both components are possibly at similar redshifts.
This result is also consistent with Hashimoto et al. (2004)
where they claimed that both eastern and western X-ray lobes 
have similar X-ray redshifts.
Based on their separation of  $\sim$ 250 kpc/h,
these results support the interpretation that 
RX J1053.7+5735 is an equal-mass cluster merger taking
place at $z$ $\sim$ 1, although 
further direct evidence for dynamical state of the cluster is 
needed for a more definitive statement about
the cluster merging state.
The six galaxies have a line-of-sight velocity 
dispersion $\Delta$$v$ $\sim$ 650 km s$^{-1}$. 
All six galaxies show clear absorption features of CaII H \& K, and
several Balmer lines, typical of early galaxies at the present
epoch, in agreement with their $I-K$ colors.
A color-magnitude diagram, constructed from
deep optical/NIR observations of the RX J1053.7+5735 field,
shows a clear red color sequence.
There is an indication that the red sequence
in RX J1053.7+5735 lies $\sim$ 0.3 to the blue of the Coma line,
qualitatively consistent with other studies investigating
other clusters at z$\sim$ 1
using optical-NIR color and pre-selection of early-type galaxies
by their morphologies. 
There is also an indication that
the fitted slope may be similar with the Coma line,
consistent with some high redshift studies,
while
contradicting studies reporting the slope evolution.
The caution has to be exercised, however, in interpreting our CM diagram,
because of  the limitation of
the data and due to the lack of morphological pre-selection.

\begin{acknowledgements}
We thank the IfA Surf's Up collaborations for their
help in acquiring the Subaru data.
JPH thanks the Alexander v. Humboldt Foundation for its
generous support.
We thank Sekyoung Yi and Sugata Kaviraj for helpful information.
We acknowledge referee's comments which improved the manuscript.
\end{acknowledgements}

\end{document}